\documentclass{article}

\usepackage{graphicx}
\usepackage[]{epsfig}
\usepackage{epstopdf}
\input{epsf.sty}

\usepackage{graphicx}
\usepackage[]{epsfig}
\usepackage{epstopdf}

\newcommand {\Cred}{}
\newcommand {\Cblu}{}

\newcommand {\bi} {\bibitem}
\newcommand {\be} {\begin{equation}}
\newcommand {\ee} {\end{equation}}
\newcommand {\bea} {\begin{eqnarray} }
\newcommand {\eea} {\nonumber \end{eqnarray}}
\newcommand {\eps} {\epsilon}

\newcommand {\lan} {\langle}
\newcommand {\ran} {\rangle}
 
\newcommand {\cA}  {{\cal A}}
\newcommand {\cC}  {{\cal C}}

\newcommand {\cG}  {{\cal G}}

\newcommand {\cR}  {{\cal R}}

\newcommand {\bc} {\begin{center}}
\newcommand {\ec} {\end{center}}
\newcommand {\bd}{\begin{displaymath}}
\newcommand {\ed}{\end{displaymath}}

\def \form#1 {eq. (\ref{#1}) }
\def \parziale#1#2  {{\partial {#1} \over \partial {#2}}}
\def \bi#1 {\typeout{#1} \item}

\begin{document}

\title{Critical behaviour of large scale dynamical  heterogeneities in glasses:\\ a complete theory\\
{\small Talk given by Giorgio Parisi at StatphysHK, Hong-Kong, July 2010} }

\author{Silvio Franz\\ {\small Laboratoire de Physique Th\'eorique et Mod\`eles
  Statistiques,} \\ {\small Universit\'e Paris-Sud 11, B\^at. 100, 91405 Orsay
  Cedex, France}\\
Giorgio  Parisi, Federico Ricci-Tersenghi, Tommaso Rizzo\\
{\small Dipartimento di Fisica, Sapienza Universit\`a di
  Roma,}\\ {\small INFN, Sezione di Roma I,  IPFC - CNR, P.le Aldo Moro 2, I-00185 Roma, Italy}}


\maketitle

\begin{abstract}
In this talk I will present a complete theory for the behaviour of large-scale dynamical heterogeneities in glasses. Following the work of \cite{I,Ia} I will show that we can write a (physically motivated) simple stochastic differential equation  that is potentially able to explain the behaviour of  large scale dynamical heterogeneities in glasses. It turns out that this behaviour is in the same universality class of the dynamics near the endpoint of a metastable phase in a disordered system, as far as reparametrization invariant quantities are concerned. Therefore Large scale dynamical heterogeneities in glasses have many points in contact with the Barkhausen noise. Numerical verifications of this theory have not yet done, but they are quite possible.
  \end{abstract}

\section{Introduction}
In this talk I will give a unified picture of the dynamic heterogeneities that arise in the slow dynamics of glassy systems. I will base my presentation on the recent  comprehensive study of the dynamics near the mode coupling transition  \cite{I}, most of it still being unpublished \cite{Ia}, that is at the basis of most of the analytic conclusions. A very important role in the arguments will be played by the results that have been obtained in \cite{CCGGPV} on the numerical simulations of glasses forming mixtures and by the extreme large scale simulations of spin glasses that have been done using the Janus machine \cite{J1,J2,J3}.

It is refreshing to see how many different pieces of the puzzle fit together and how this new picture helps us to answer to old questions and to formulate new questions. As we shall see many progresses can be done if we concentrate our attention on reparametrization invariant quantities, i.e. on the relations among quantities that are obtained by eliminating the time parametrically. We present here a complete theoretical approach that is able to predict many of the observed quantities.

\section{Let us eliminate the time from the plots!}
\subsection{The importance of being reparametrization invariant}
The very strong slowing down in glassy material is a very impressive feature. 
Often most of the attention  is devoted to the study of the  time dependence of the observables. 
However this may be not wise. 

It is important to understand first {\em what} happens and to postpone the study of {\em when} it happens. It would be very difficult to understand {\em when}, if we do not know {\em what}. In plain words let us suppose  that we have a movie of the evolution of a glass. We would like to have an understanding of what we see on the various pictures without having to know the exact time at which each picture is taken. We stress the importance of having a well-characterized synchronic description before attempting a diachronic one \footnote{If an adiabatic approximation can be done, the actual time dependence is irrelevant for computing the evolution of a state.}. From this point of view we can get  a greater insight if we consider two observables, e.g. $A(t)$ and $C(t)$ and we plots $A(t)$ versus $C(t)$, parametrically in time. Sometimes we say that the corresponding function $\cA(C)$ is reparametrization invariant, because it would not change if we redefine the time in a diffe!
 rent way.

When the dynamics become very slow, reparametrization invariant quantities are much simpler to be understood theoretically and they have a much higher degree of universality. Indeed  when the dynamics become very slow the actual time dependence becomes irrelevant in deducing the relation among different quantities: we can do adiabatic-like approximations where the systems goes through quasi-equilibrium states. The actual time dependence is much harder to find out: for example in the case of a system where the dynamics is dominated by barriers crossing, the actual time dependence is related to the distribution of the barriers that is a rather complex quantity to compute. 

Of course in a dully world, where there is only one quantity that we can measure at a given time, reparametrization invariant quantities would be trivial. Fortunately there are many observations that can be done at the same time, and we have many reparametrization invariant quantities to study. In spite of the fact that reparametrization invariant quantities have already been noticed in the past, their importance is not widely appreciated. We will present now some examples that illustrate their importance and simplicity.

Of course we want (and eventually we will be able) to understand the time dependence of various quantities, but this very difficult task may be successful only after that we establish a firm command of the behaviour of reparametrization invariant quantities.

\subsection{Fluctuation dissipation relations in off-equilibrium systems.}

Fluctuations dissipations relations in off-equilibrium systems are one of the most powerful tools to investigate the properties of the energy landscape of glassy systems. They are well known and I will just recall the definitions and the main results \cite{CUKU,FM}.

During aging we define a two times correlation function $C(t_w,t_w+t)$ and a relaxation function $R(t_w,t_w+t)$. For a given $t_w$ one defines a function $\cR_{t_w}(C)$ by eliminating parametrically the time:

\be
\cR_{t_w}(C(t_w,t_w+t))=R(t_w,t_w+t)
\ee

The fluctuation dissipation relations tell us that the limit 
\be
\cR(C)=\lim_{t_w \to \infty} \cR_{t_w} (C)
\ee
is well defined. Moreover the function $\cR(C)$ (that is experimentally observable) encodes interesting properties of the (free-)energy landscape. In particular it is know that:
\begin{itemize}
\item The form the function  $\cR(C)$ can be deduced by general arguments: for example in structural glasses we expect that 
$\cR(C)$ has slope $1$ for $C>C^*$ and $\cR(C)$ has slope $m$ for $C<C^*$, where $C^*$ is the position or the plateau in the two times correlation function $C(t_w,t_w+t)$.
\item The function $\cR(C)$ does not change if the same system is put in an  off-equilibrium situations using a method that is different from a temperature jump (as done in aging): e.g. it is subject to a small time dependent force, for example the very gentle stirring of a fluid. 
\item Even in infinite range models, where all static quantities can be computed analytically, also the function  $\cR(C)$ can be computed analytically in spite of the fact that (at the present moment) we are unable to compute $C(t_w,t_w+t)$ for large $t_w$.
\end{itemize}

It is crucial to note that it is possible to relate the function  $\cR(C)$ to a purely equilibrium quantity (i.e. the probability distribution of the overlaps $P(q)$). Indeed this is the main point of this talk: it ought to be possible to compute reparametrization invariant quantities without having to solve the dynamical equations, but computing the properties of a simple model (in the best of the possible worlds an equilibrium model).

In a similar way we can introduce local fluctuation-dissipation relations that have interesting properties that cannot be described here for lack of space
\cite{LOCAL}.

\subsection{Dynamical heterogeneities in mean field theory}\label{DHMFT}
We consider a very simple case: a mean field model \cite{CAV,MP, KWT, KT,REP} (i.e. a random $p$-spin Ising model on the Bethe lattice) that has a standard dynamical transition at a temperature $T_d$. 
The correlation time of a system of $N$ spins diverges when $N\to\infty$ and $T\to T_d$. 
For a given system we can define the correlation functions as
\begin{equation}
C_{N,T}(t)=\overline{q(t',t'+t)}\ ; \ \ C_{N,T}^2(t)=\overline{q^2(t',t'+t)}\, ,
\end{equation}
where the overline denotes the average over $t'$ and $q(t',t'+t)$ is the (normalized) overlap between a configuration at time $t$ and another configuration at time $t'$.

The quantity  $C_{N,T}^2(t)-(C_{N,T}(t))^2$ tells us how strongly the correlations fluctuates in time: this is the simplest example of dynamical heterogeneity.

The behaviour of the function $C_{N,T}(t)$ is rather complex:
\begin{itemize}
\item For $T>T_d$ the function $C_{T}(t)\equiv \lim_{N\to\infty}C_{N,T}(t)$ can be studied using mode-coupling type equations. Exactly at $T_d$ for large $t$ we have that  $C_{T_d}(t) \approx C^* +B t^{-b}$ (for a review see for example \cite{KOB}) . 
\item Exactly at $T_d$ the function $C_{T_d,N}(t)$ has a plateau at  $C^*$ for large $N$ and it decrease to zero when the time goes to infinity slower and slower when $N$ goes to infinity.
\item In the mean field case (where the mode coupling equations are correct) one can distinguish a {\em short time }$\beta$ regime where the $C\approx C^*+Bt^{-b}$ and a later {\em large time}  regime $\alpha$ regime where $C\approx C^*-At^a+O(t^{2a})$. The $\alpha$ regime disappear (in mean field theory) below the dynamical temperature.
\end{itemize}

We stress that at the present moment we have a very little command on the form of the function $C_{N,T}(t)$, we only know the value $C^*$. However in the critical region it is possible to make detailed predictions on the behaviour of the function $\cC^2_{N,T}(C)$ defined as
\be
\cC^2_{N,T}(C_{N,T}(t))=C^2_{N,T}(t) \ .
\ee
These predictions will be described in details in \cite{Ia}. Let me present the simplest and more surprising one (coming from numerical analysis or from the analytic study of the differential equation (\ref{MOCK})).
If we define
\be
\lim_{N\to\infty} \cC^2_{N,T_d}(C)= \cC^2_{d}(C) \, ,
\ee
we find
\be
\cC^2_{d}(C)= {CC^*} \ (C<C^*)  \ ;\ \ \ \cC^2_{d}(C)=C^2 \ (C>C^*)\  . \label{MIRA}
\ee

If we define the dynamical susceptibility as $\chi(C)=\cC^2_{d}(C)-C^2$ we find:
\be
\chi(C)= {C(C^*-C)} \ (C<C^*)  \ ;\ \ \ \chi(C)=0 \ (C>C^*)\  . \label{MIRA1}
\ee
The corresponding Binder cumulant can be computed and it is given by
\be
B(C)=3-{(C^*)^2\over 2C(C^*-C)}
\ee
At the point where $\chi(C)$ is maximum we have  $C=C^*/2$ and $B=1$.

The previous equations have a very simple meaning: in the long time limit for large systems at $T_d$ the quantity $C(t',t'+t)$ may have (with probability one) two values (0 or $C^*)$. 
The result is very simple in spite of the rather complex dependence of the correlation function on time and $N$.

\begin{figure}[t!]
\includegraphics[width=.6\columnwidth]{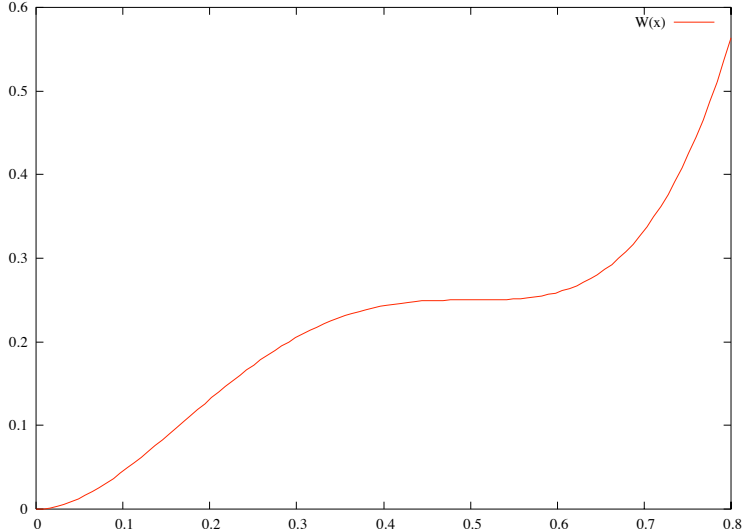}
\caption{ A schematic view of the potential $W(q)$ computed at the dynamical temperature.}
\label{SHAPE}
\end{figure}

It is interesting that we can obtain the results for the function $\cC^2_{d}(C)$ by considering the following differential equation,  where we have introduced a fictitious time $s$:
\be
{dC_\eta(s)\over ds}= -W'(C_\eta(s))+\eta \label{MOCK}\, ,
\ee
with the boundary condition $C_\eta(0)=1$, where  $W(C)$ has the shape given in fig. (\ref{SHAPE}), i.e.  minimum at the origin and an horizontal flex at $C^*$ and $\eta$ is a random number whose variance goes to zero as a power of $N$. If we are interested to compute only  the function $\cC^2_{d}(C)$, the precise $N$ dependence of $\eta$ is not relevant.

We define
\be
C(s)=\overline{C_\eta(s)}\, ,\ \ \ \ C^2_d(s)=\overline{C_\eta^2(s)}\;,
\ee
where $\overline{\;\cdot\;}$ means average over the noise $\eta$.
If we eliminate the fictitious time $s$ in a parametric way, we find the result in equation (\ref{MIRA}).

The learned reader may be quite surprised: equation  (\ref{MOCK}) is quite different from a mode coupling equation and it does not have all the complications of the mode coupling equation (memory kernel, non-locality in time and all other paraphernalia). This is the main point we are making in this talk: as far as reparametrization invariant quantities are concerned, the two equations are equivalent, in spite of the enormous difference in complexity. Equation  (\ref{MOCK}) is enough to get a physical understanding of the behaviour of the fluctuations.

\subsection{Two approaches to finite volume analysis}

We would like to spend a few words on  what happens in a finite volume. In the infinite volume limit the correlation $C(t',t'+t)$ does not depends on $t'$, i.e. it is an intensive quantity that has fluctuations proportional to $1/V$. When we go to finite volume systems, the fluctuations may be  very strong. 

We would to point out that in a finite volume there are two different procedures to make the averages:
\begin{itemize}
\item The procedure that has been described previously: we compute the averages over $t'$ of the observables $A(t',t'+t)$ and of the correlation $C(t','t+t)$ and we plot parametrically in time $A(t)$ versus $C(t)$.
\item For each $t'$ we compute
\be
\cA(t',C)\equiv{\int dt \delta(C(t',t'+t)-C)A(t',t'+t) \over \int dt \delta(C(t',t'+t)-C)}\, ,
\ee
At the end we do the average over $t'$ of $\cA(t',C)$ in order to get $\cA(C)$.
\end{itemize}
There are advantages and disadvantages in both procedures. It is important to stress that the results may be quite different depending on the chosen procedure. It is likely that in some cases it would be convenient to analyze the data using both approaches.

\section{Dynamical heterogeneities in short range models: a naive approach}
\subsection{Overlaps and correlation function}
Let us consider a system where the approach at equilibrium is very slow and the correlation functions takes a very long time to decay. The correlation functions have a plateau that may become very long. Quite often the parameter that controls the length of the plateau is the temperature (near a dynamical critical point) or the waiting time.

Given two configurations that we denote ($\sigma$ and $\tau$), we define
$ q_{\sigma,\tau}(x)$
as the similarity (overlap) of the two
configurations in the region of space around the point $x$. Usually the definition is such that

\begin{itemize}
\item $q_{\sigma,\tau}(x)=1$   for
identical configurations,
\item $q_{\sigma,\tau}(x)\approx 0$  for uncorrelated
configurations. 
\end{itemize}

In the spin case $ q_{\sigma,\tau}(x)$ is just the product of the spins of two configurations at point $x$. For fluids different definitions are possible:
we can take
$q_{\sigma,\tau}(x)=1$, if a region around $x$ of size
$a$   has the same particle content in the
configurations $\sigma$ and $\tau$; otherwise,
if the particle content is different, $q_{\sigma,\tau}(x)=0$.
In the simplest case, where there is one kind of particles we can just take
$
 q(x)=\sigma(x)\tau(x),
 $
 where
$\sigma(x)$ and $\tau(x)$ are the smeared densities around the point $x$.

In the aging regime the correlation function $ C(t_w,t_w+t)$ is given by 
\be
C(t_w,t_w+t)=\overline{ q(t_w,t_w+t;y)}\, ,
\ee
where the overline denotes the space average over the point $y$ and we have used the shorthand notation:
\be
q(t_w,t_w+t;y)\equiv q_{\sigma(t_w),\sigma(t_w+t)}(y)\ .
\ee

We are interested in computing the average of the connected correlation functions
\be
G(t_w,t_w+t;x)=\overline{q(t_w,t_w+t;y) q(t_w,t_w+t;y+x)}- C(t_w,t_w+t)^2
\ee
These correlations can be used to trace the properties of heterogeneities in the dynamics \cite{DAS,SH,PL,srpsp}.

Physically the  strong slowing down observed in glasses is connected to the existence of cooperatively rearranging regions. These regions manifest themselves as  large heterogeneities in the dynamics that imply the existence of long range correlations in the function $G(t_w,t_w+t;x)$.

Our aim is to get precise predictions on the form of the correlations,  on the behaviour of the correlations length, on critical exponents, upper critical dimensions and all that.
\begin{figure}[t!]
\includegraphics[width=.9\columnwidth]{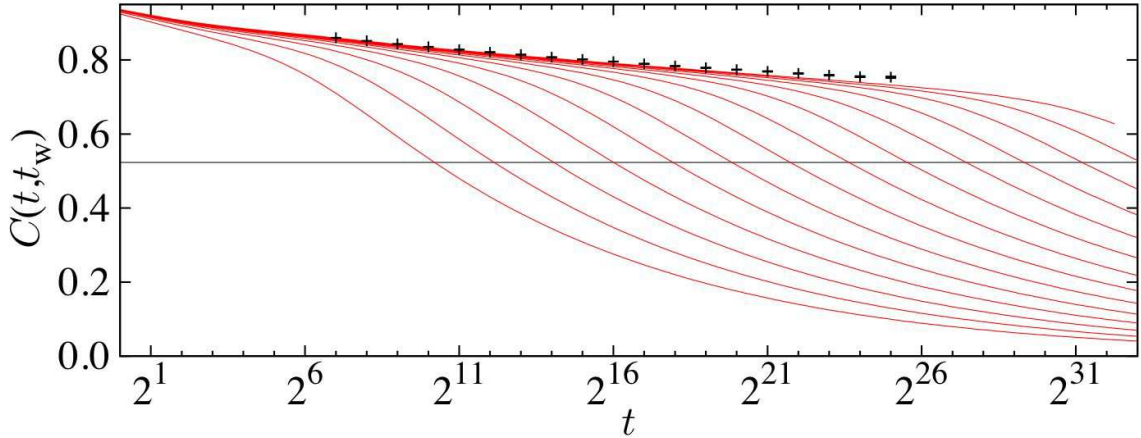}
\caption{ The correlation $C(t_w,t_w+t)$ in spin glasses below $T_c$ during aging: the different curves are taken at different times $t_w$ (data taken from \cite{J1}). The crosses indicate the extrapolation at infinite waiting time $t_w$, The horizontal line is the asymptotic position of the plateau (estimated using other information). }
\label{JANUS}
\end{figure}
 
According to the previous discussions we eliminate parametrically the time $t$ (the new clock becomes $C$): we consider  the correlation function
\be
\cG(t_w,x|C)
\ee
Our aim is to characterize in some way the expectation values of local quantities in the ensemble where $t_w$ goes to infinite at fixed $C$. It may be convenient to indicate the average as
$
\lan \cdot \ran^{D}_C\ .
$

\subsection{A bold conjecture}

Let us do something completely different: we take two replicas $\sigma$ and $\tau$. The first one is a standard equilibrium configuration while the second is a configuration constrained to have an overlap $q$ with the first one.
We  restrict the statistical sum to all the configurations $\tau$ that have an overlap $q$  with $\sigma$.
In other words we define the following expectation values:

\be \lan A \ran^R_{q}={\int d\tau \exp(-\beta H(\tau)) \Cred{\delta (q-q(\sigma,\tau))} A(\tau) \over
 \int d\tau \exp(-\beta H(\tau))\Cred{ \delta (q-q(\sigma,\tau))}}\, , \label{PART}
\ee
where the average over $\sigma$ is implicit and the subscript $R$ stands for replica.

At this stage it is convenient to introduce the so called replica potential $W(q)$ \cite{REP,CPR,FP,PS,PZ}.
We consider an equilibrium configuration $\sigma$. We consider a restricted partition function $Z_\sigma(q)$ where we sum over all the configurations that are at an overlap $q$ with the configuration $\sigma$;

\be
{Z_\sigma(q)\over Z}= {\int d\tau \exp(-\beta H(\tau)) \Cred{\delta (q-q(\sigma,\tau))}  \over
 \int d\tau \exp(-\beta H(\tau))}=
\exp( -N W_\sigma(q)) \ .
\ee
The restricted partition function $Z_\sigma(q)$ is the denominator in equation (\ref{PART}). Finally
\be
W(q)=\overline{W_\sigma(q)}= \overline{ -\log(Z_\sigma(q))\over N} \, ,\ee
where the overline  here denotes the average over the different equilibrium configurations $\sigma$.

Therefore $W(q)$ is $\beta$ times the increase in free energy when we restrict the sum to those configurations that have an overlap $q$ with a generic configuration $\sigma$.  In other word
  \Cred{$ \exp( -N W(q))$ is the typical probability of finding a configuration with overlap $q$.}
  
  The naive conjecture is that in the region of $C$ that corresponds to large times (i.e. those where ${dC\over dt}$ is very small), the two previously defined ensemble do coincide for $q=C$:
  \be
  \lan A \ran^{D}_C= \lan A \ran^R_C \ \label{MAGIC} \, ,
  \ee
  where in the first case $A$ is the product of factors of the form $q(t_w,t_w+t;x)$, while in the second case it is the product  of factors of the form $q_{\sigma,\tau}(x)$.

This is a far reaching conjecture whose validity should strongly simplify the theoretical analysis and the interpretation of the data. It tames all the difficulties of the dynamics and it reduces the analysis to an equilibrium computation.
\begin{figure}[t!]
\includegraphics[angle=270,width=.46\columnwidth]{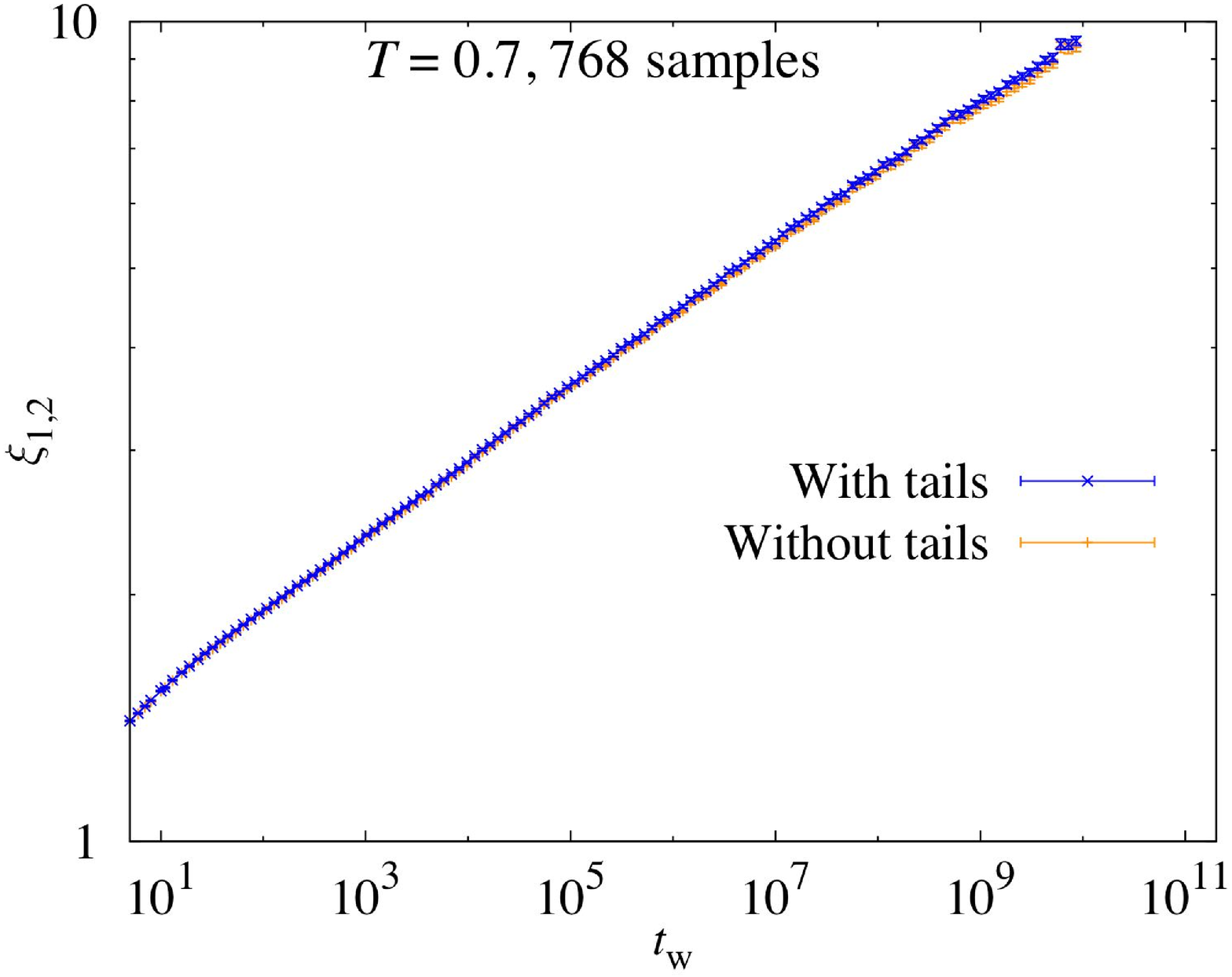}\includegraphics[angle=270,width=.51\columnwidth]{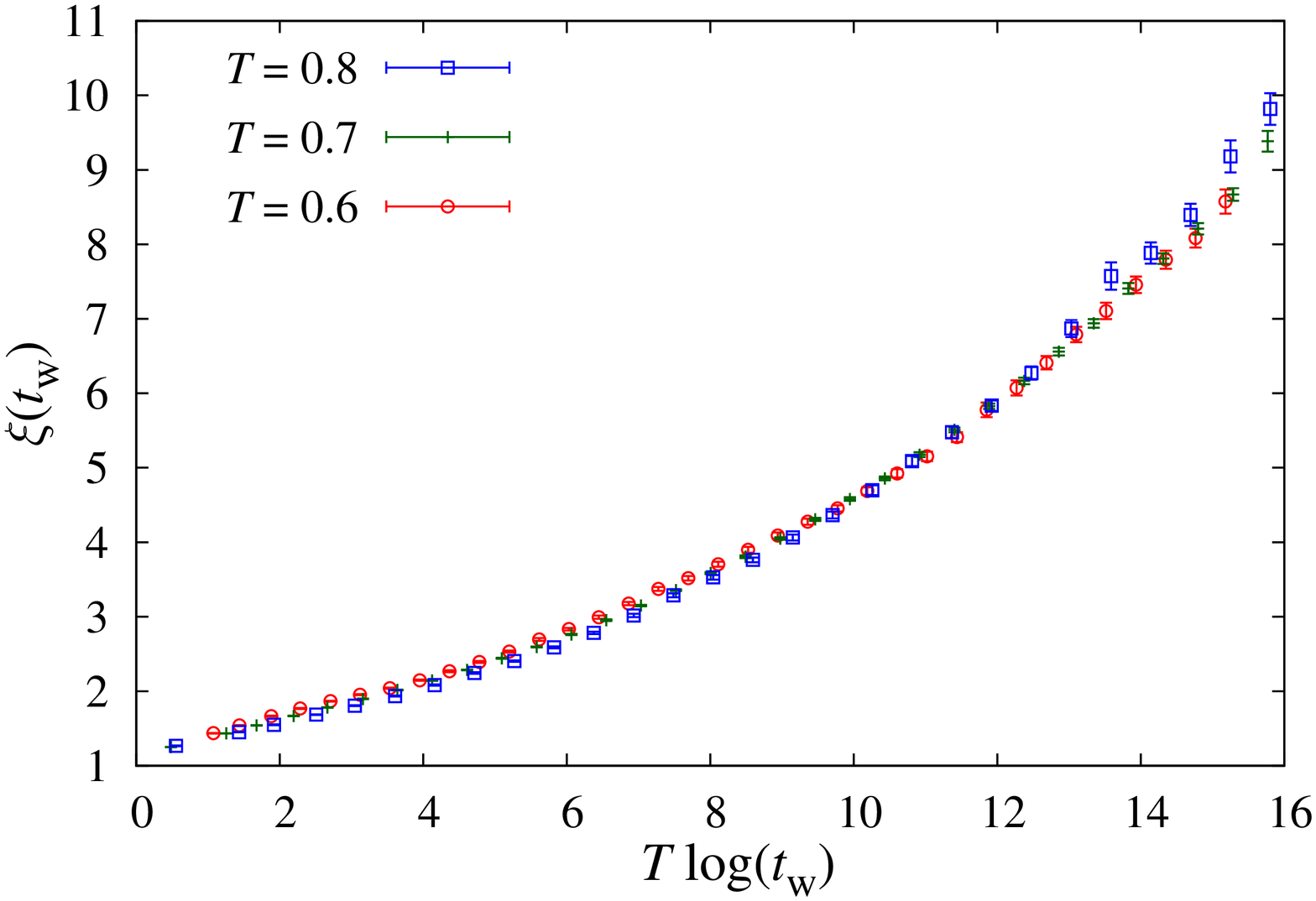}
\caption{Left panel: the correlation length in spin glasses as a function of time at $0.6 T_c$; right panel: the correlation length as a function of the $T\log(t)$ at various temperatures (data taken from \cite{J1}.  }
\label{XI}
\end{figure}

\subsection{A Favorable case: spin glasses}
We consider the case of a three dimensional spin glass at zero magnetic field. In this case the replica potential $W(q)$ is symmetric ($W(q)=W(-q)$) and it satisfies the following relations:
\begin{itemize}
\item For $q<q_{EA}$, $W(q)=0$.
\item For  $q>q_{EA}$, $W(q)>0$, where  for small $q-q_{EA}$ we should have  $W(q)\propto (q-q_{EA})^\delta$, with $\delta\approx 7-8$.
\end{itemize}

It is interesting to look to reparametrization invariant quantities. 
In this case it should be possible to do an explicit check of the correctness of equation (\ref{MAGIC}). Unfortunately the direct comparison is not so simple because both the $t_w$ dependence in the dynamics and the dependence on the size of the box $L$ in the equilibrium data are strong and the extrapolations to infinite $t_w$ and infinite $L$ are not trivial. We can use the time-space equivalence principle \cite{J3} that state that the finite volume corrections and the finite time corrections are similar for a box of size $L^*(t)$:
\be
 \lan q(x) q(0)\ran^{D}_q  \approx \left. \lan q(x) q(0)\ran^R_q \right|_{L^*(t_w)} \label{MAGIC1} \ .
\ee

But how to determine $L^*(t)$? We could do a fit in order to match the correlations functions, however we can proceed in a niftier way. 

In the spin glass case quenched disorder is present: we can study the correlations induced by disorder. This can be done by  simulating the dynamics starting from the two different configurations. We can consider the overlaps among two different configurations both at time $t_w$, i.e. $q(\sigma_1(t_w),\sigma_2(t_w),x)$ and define their correlation function $C_4(t_w,x)$. According to general principles we should have that
\be
\lim_{t_w \to \infty}C_4(t_w,x) = \lan q(x) q(0)\ran^R_0 \ .
\ee
The space-time equivalence principle implies that
\be
C_4(t_w,x) \approx \left. \lan q(x) q(0)\ran^R_0 \right|_{L^*(t_w)} \label{MAGIC2}
\ee
where $\left. \lan \cdot \ran^R_0 \right|_L$ denotes the expectation values computed in a box of size $L$.
\begin{figure}[t!]
\includegraphics[angle=270,width=.46\columnwidth]{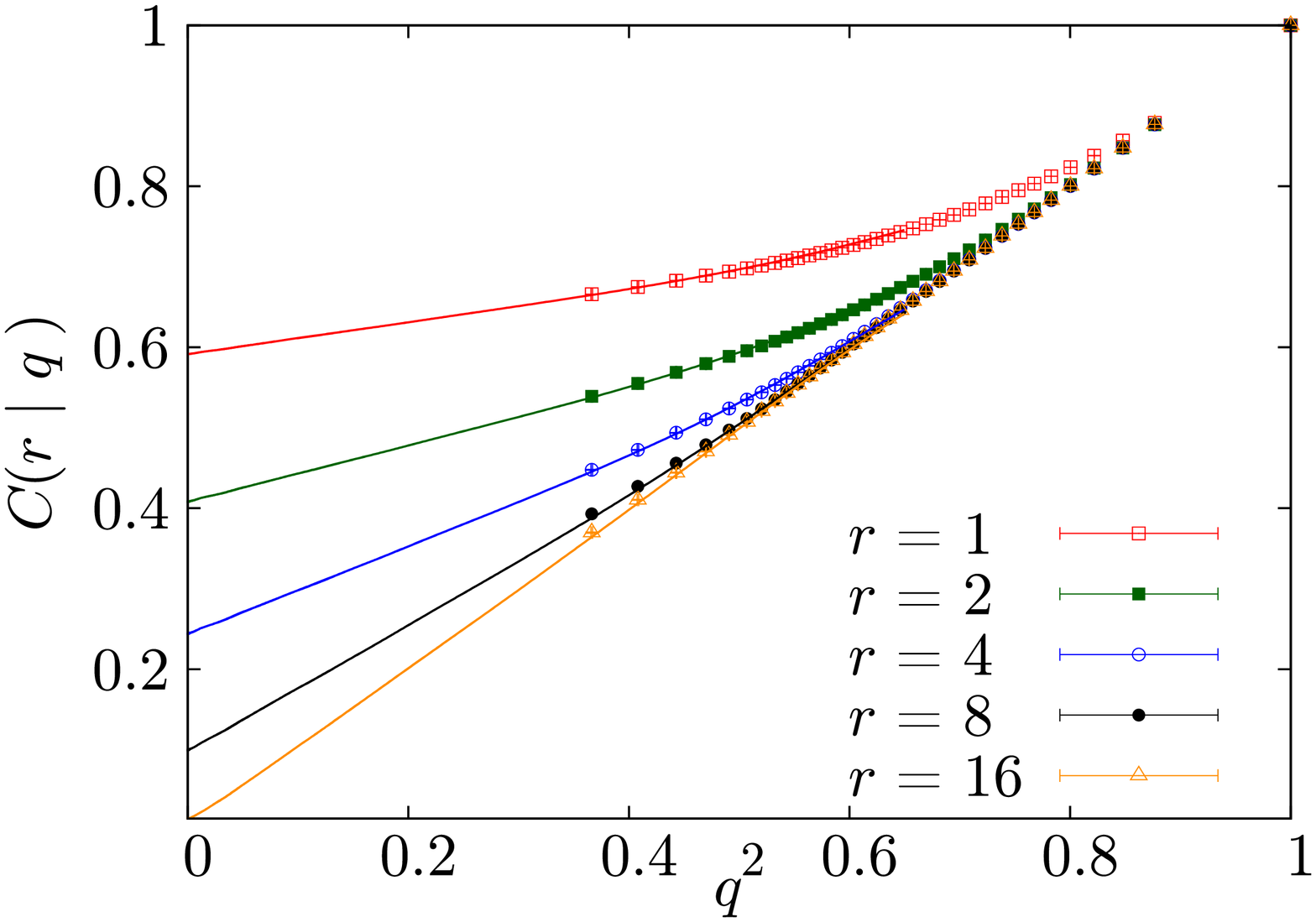}\includegraphics[angle=270,width=.51\columnwidth]{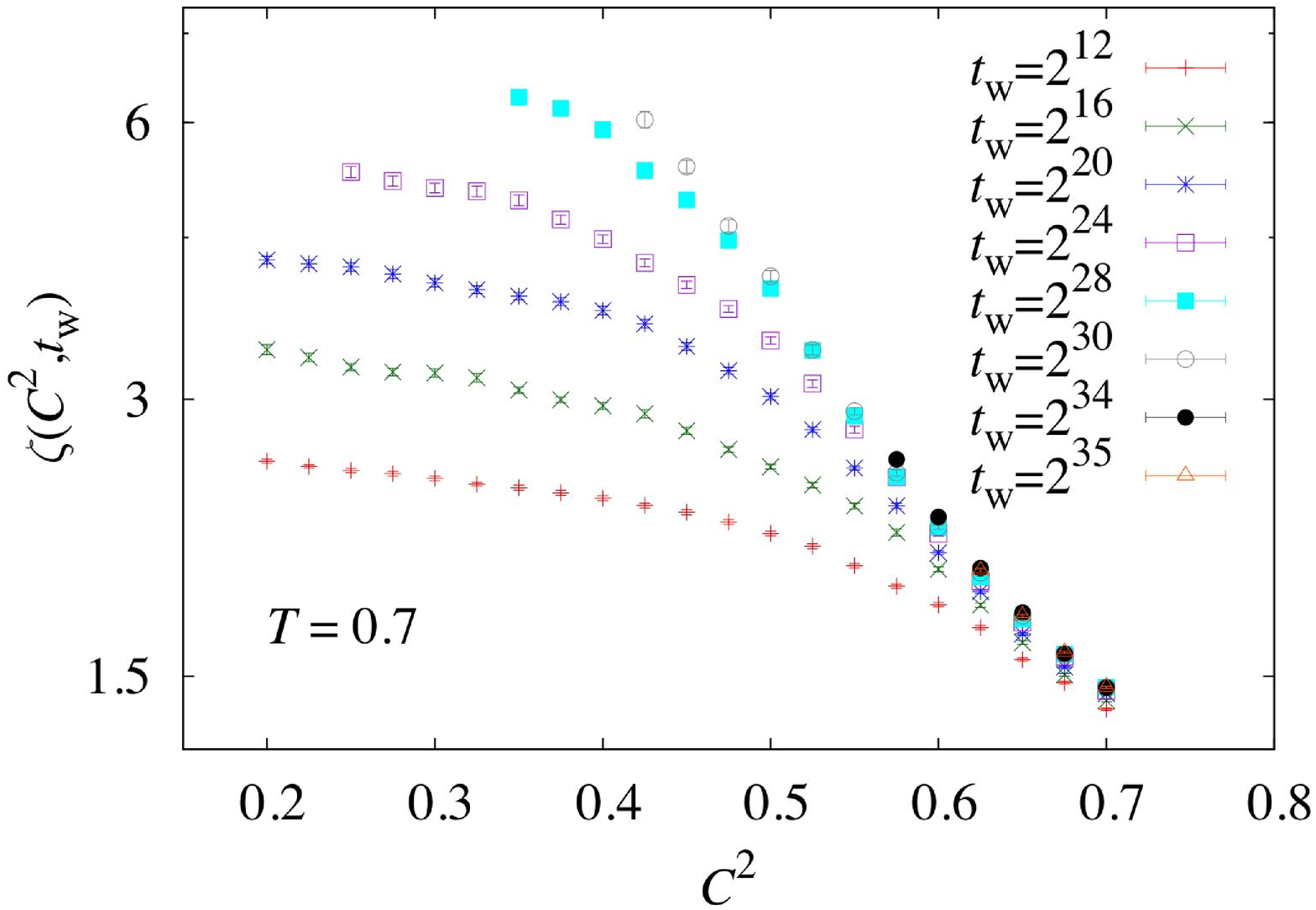}
\caption{Left panel: the correlation function at distance $r$ as function of $q^2$ or $C^2$, full line equilibrium data, points dynamical data \cite{J3}; right panel the correlation length for heterogeneities $\zeta$, from the dynamical data as  function $C^2$ at different $t_w$ \cite{J1}.  }
\label{ZETA}
\end{figure}

One finds in the low temperature phase (i.e. at a temperature $T \approx 0.6T_c)$:
\begin{itemize}
\item Also in the infinite time limit  the function $C_4(t_w,x)$ goes to zero: $\lim_{t_w \to \infty}C_4(t_w,x)\propto x^{-\alpha}$ with $\alpha \approx 0.4$ for large $x$.
\item At finite time the function $C_4(t_w,x)$ goes to zero at large distances exponentially (or faster) and one can define a correlations length $\xi(t)$.\footnote{The correlation length $\xi(t)$ increases  as $ t^{\lambda(T)}$ with $\lambda(T)$ linear in the temperature. This very interesting and not well understood phenomenon involves non-reparametrization invariant quantities and its discussion would be out of place here.} The correlation length as a  function of time is shown in fig. (\ref{XI}).
\item  It is possible to satisfy equation (\ref{MAGIC2})  with an $L^*(t_w)\approx 3.7 \xi(t_w)$.
\end{itemize}
Using these results one can just determine the correlation length $\xi(t_w)$ from the $C_4(t_w,x)$ correlation and check directly equation (\ref{MAGIC1}). The agreement is excellent, see fig. (\ref{ZETA}), left panel. If we plot the correlation length for the dynamic heterogeneities we find the results in the right panel of fig. (\ref{ZETA}). One can see that in the $\beta$-region (i.e. high $C^2$) the correlation length becomes independent from $t_w$ while in the $\alpha$-region (i.e. low $C^2$) the correlations length increases with $t_w$.

The interested reader should look to the original papers in order to know what we have learned on the behaviour of dynamical heterogeneities in this case. Here we want to stress that the naive approach to the computation of the heterogeneities works very well; this is likely due to the vanishing of the replica potential $W(q)$; as we shall see later, in structural glasses the situation is more complex.

\section{Near the dynamical transition}

\subsection{The behaviour of the replica potential}
We have introduced the replica potential $W(q)$ in the previous section. In mean field models near the dynamical transition a secondary minimum appears \cite{REP}.
The value of $W(q^*)$ at the secondary minimum is the configurational entropy $\Sigma_C$.
  Indeed the probability of a given valley is $\exp(-N\, W(q^*))$ and their number is  $\exp(N\,W(q^*))$.

\begin{figure}[t!]
\includegraphics[angle=0,width=.4\columnwidth]{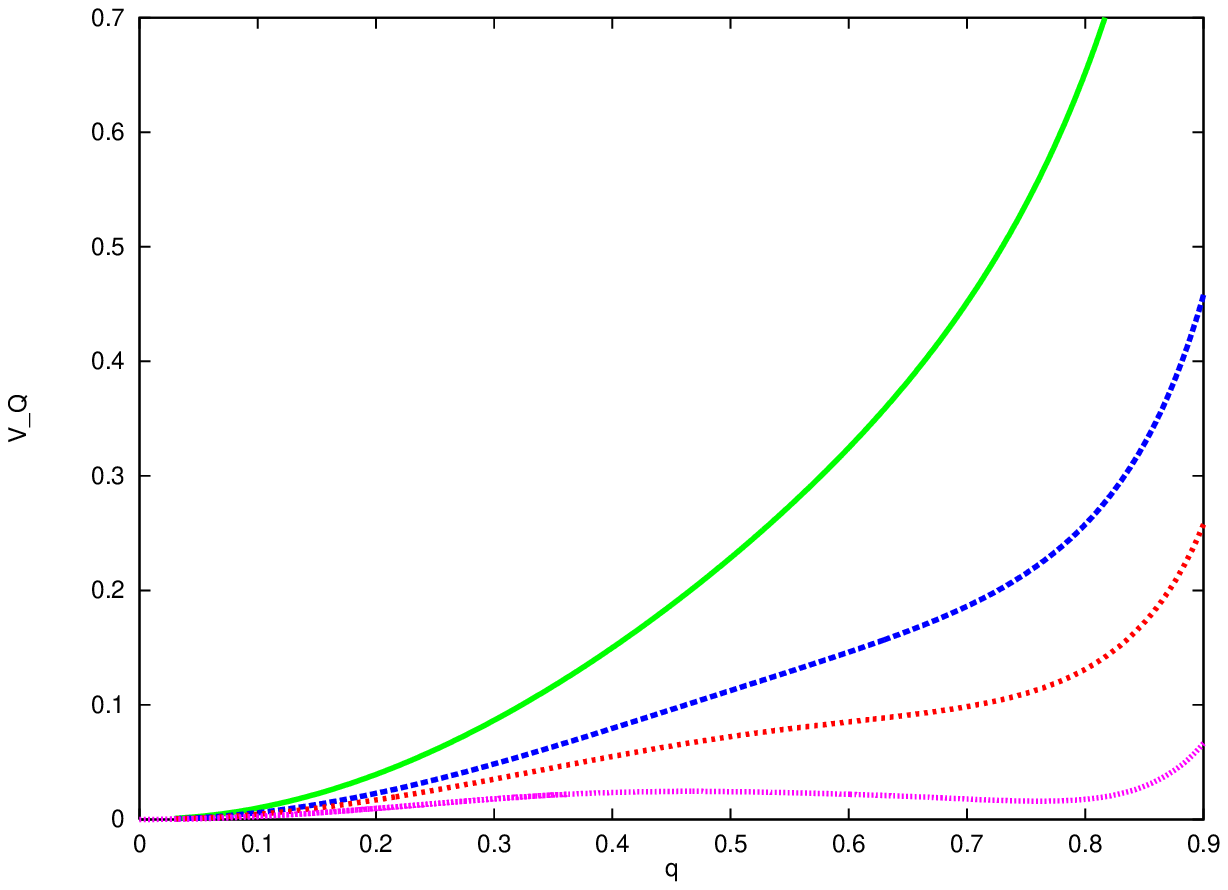}  \includegraphics[ width=.5\columnwidth]{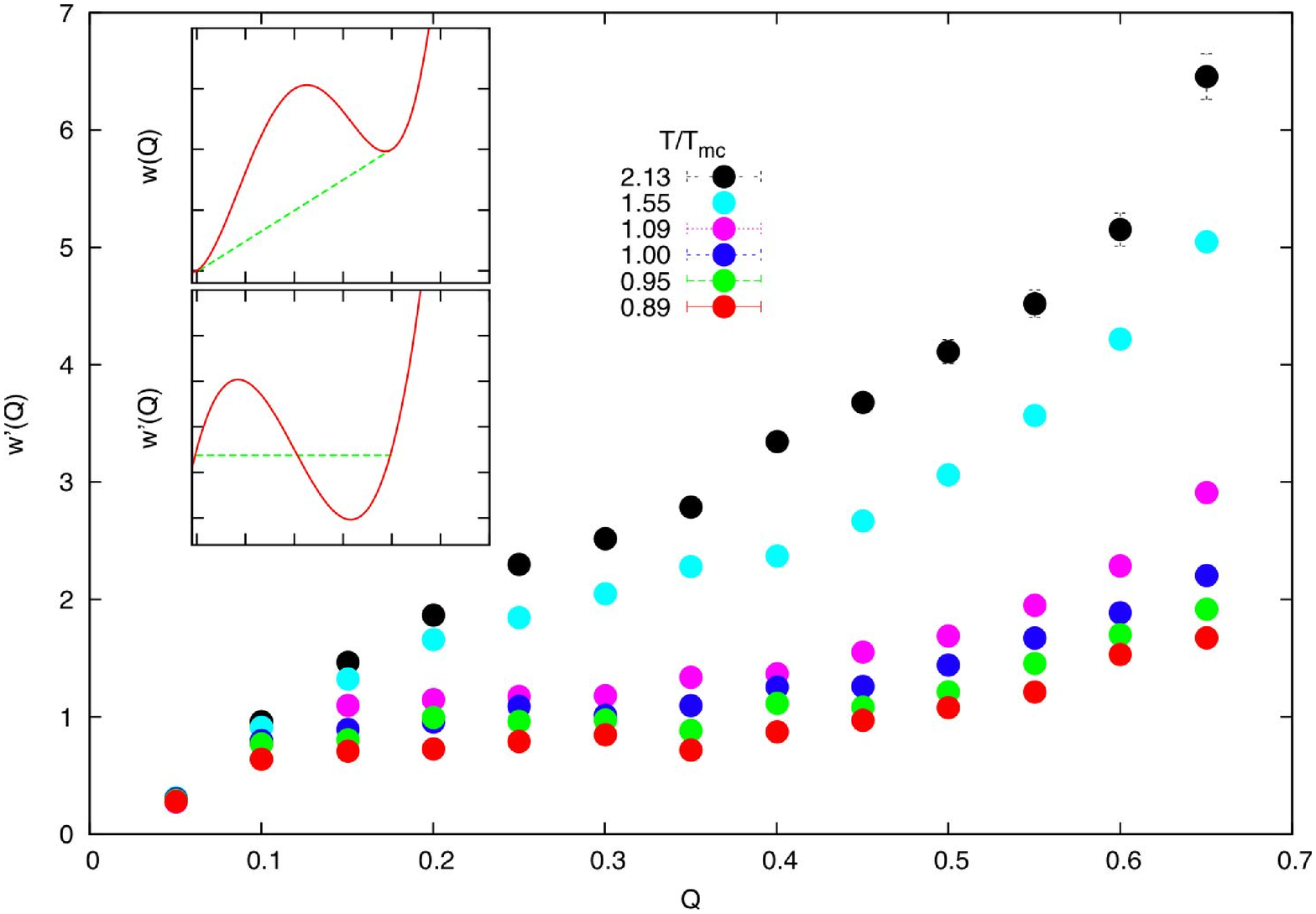}
\caption{ Left panel: the value of the potential $W'(q)$ for different temperatures, decreasing from above to below in the mean field approximation. Right panel: numerical evidence of the Maxwell phenomenon for soft spheres new the dynamical transition \cite{CCGGPV}}
\label{WM}
\end{figure}

 In the dynamics we expect naively that there is slowing down in the correlations near the dynamic transition, as discussed in details in section (\ref{DHMFT}). Moreover we should have that for $T<T_d$, the equilibrium correlation function does not decays to zero, i.e. $\lim_{t\to \infty} C(t)=C^*\ne 0$. However in short range model the convexity of the free energy implies that the correct formula of the potential is given by the convex envelope of $W(q)$, see the insert in fig. (\ref{WM}). It is easy to check that the Maxwell construction implies a constant $W'(q)$ in a wide region, that correspond to phase separation in configuration space.
 
 Numerical simulations near the dynamical point for a binary mixture of three dimensional soft sphere clearly show the presence of a Maxwell phenomenon: this is an important test of the basic assumptions of the theory.

One finds that there are two different regimes:
\begin{itemize}
\item The $\beta$-regime, $C(t)>C^*$. We are approaching the saddle point. Heterogeneities are due to fluctuations in the shape of the barrier.
This is a quasi-equilibrium situation. The Dynamics gives the same results of the constrained replica approach and the equivalence should be perfect quite near to the plateau.
\item The $\alpha$ regime, $C(t)<C^*$. Heterogeneities are due to variations in the {\Cblu escape} time due to fluctuations in the shape of the barrier.
Dynamics does not give the same results of the constrained replica approach: there is no Maxwell construction in the dynamics, therefore it seems that we do not have any way to use the dynamic-replica connection in this region.
\end {itemize}

It is interesting to note that the mode coupling equations  for the frozen correlations in the region $T<T_d$   have been derived in a purely equilibrium setting in a recent remarkable paper \cite{REM}, adding more evidence for the dynamic-replica connection in the $\beta$-regime.

\subsection{A world of caution: two different correlation functions for dynamical heterogeneities}

It is usually not recognized that we can define two different correlation functions for studying the behaviour of dynamical heterogeneities. They display a quite different behaviour.
 In order to be clear we denote by $\sigma$ the configuration at time $t'$ (the initial configuration) and by $\tau$ the configuration at time $t'+t$. We also indicate by an overline the average of the $\tau$ configurations and by $\lan \cdot \ran_\sigma$ the average over the $\tau$ configurations at fixed $\sigma$ (we assume that the dynamics is stochastic, as it happens for the Langevin equation or for Monte Carlo).

The standard correlation function is 
\be
G_0(x)= \overline{\left\lan q_{\sigma,\tau}(x)q_{\sigma,\tau}(0) \right\ran_{\sigma} }- 
\overline{ \left\lan q_{\sigma,\tau}(x) \right\ran_{\sigma}}\;
\overline{ \left\lan q_{\sigma,\tau}(0) \right\ran_{\sigma}} \ .
\ee
However we can define a different correlation function
\be
G_1(x)=\overline{\left\lan q_{\sigma,\tau}(x)q_{\sigma,\tau}(0) \right\ran_{\sigma} -  \left\lan  q_{\sigma,\tau}(x)\right\ran_{\sigma}\left\lan q_{\sigma,\tau}(0)  \right\ran_{\sigma}}\ .
\ee
The correlation function $G_1(x)$ measures the connected correlations of $\tau$ at fixed $\sigma$, i.e. the heterogeneities induced by the noise, while the correlation function $G_0(x)$ takes care also of the correlations induced by the fluctuations in the initial configuration.

Neglecting loops  near the critical point the theoretical prediction (in momentum space) are \cite{CPR, BB0, BB1,Ia}:
\be
\widetilde G_0(k,\eps)\approx{A\over (\xi^{-2}+k^2)^2}+{B\over \xi^{-2}+k^2}\ , \ \ 
\widetilde G_1(k,\eps) \approx{C\over \xi^{-2}+k^2}\ . \label{PROP}
\ee
where $\xi$ is the correlation length.
Unfortunately there are very few measurements done for the second correlation function $G_1$, but, especially in the theoretical analysis, one has to take care of its existence. It is also important to note the presence of a double pole in $G_0$ and not in $G_1$.

As we shall see later the situation is very similar to that of a ferromagnetic material in presence of a random field. In this case one can define two correlation functions that are characterized by different exponents \cite{PAS1,PH}.

\subsection{A proof of the replica-dynamics connection in perturbation theory}
Here we present the hard analytic results that form the backbone of this talk. We can now stop to wave hands and to make analogical argument and we can start to look to the bare formulae. Here I will only present the results; the actual derivations can be found in \cite{I,Ia}. These analytic results concern mainly the $\beta$ regime and not the late decay in the $\alpha$ regime.

We will consider for simplicity a short range $p$-spin model (or any other model with a simple mean field theory). It is crucial to note that we can construct two quite different perturbative expansions:
\begin{itemize}
\item The perturbative expansion in the dynamics: the starting point (i.e. the zero-order approximation)  is the solution of the mode coupling equations in the infinite range model. This perturbation can be constructed for example by using the Martin-Siggia-Rose formalism \cite{SBBB,Ia}.
\item We can study the model with one replica constrained to stay at a fixed overlap from a reference configuration. In this case we can just construct the standard loop expansion for the two coupled replicas in the framework of the replica formalism. Essentially we work with an action of an order parameter matrix $q_{ab}$ with $a,b=1,\dots n$ in the limit of $n=1$ replicas. The general diagrammatic rules of \cite{DKT} can be considerably simplified due to the $n=1$ limit and display a great similarity to the Random Field Ising Model \cite{PAS1}.
\end{itemize}

At all order in perturbation theory in the loop expansion one can prove that the leading terms near $C^*$ do coincide. More precisely in the limit $T\to T_d$ and $C$ near  to $C^*$ (and bigger than $C^*$), the two theories only differ by subleading terms. Therefore in perturbation theory  they have the same critical exponents. 

There are many more detailed results; the main ones are:

\begin{itemize}
\item The correlations (neglecting loops) are given by the previous expressions (\ref{PROP}).
\item The upper critical dimension is 8: the previous expressions   (\ref{PROP}) for the correlations should be valid for dimensions greater than 8.
\item The theory can be related to a stochastic differential equation that has a clear physical meaning (we will explore this point later on). This equivalence can be shown diagrammatically at all orders and also in a non-perturbative way similar to \cite{CARDY}. 
\item Perturbative dimensional reduction is valid, i.e. the theory is equivalent to  a non-random theory in dimensions $D-2$.
\item The sum of the perturbative expansion is ill-defined ({\em sic erat in fatis}). Perturbation theory indeed misses the existence of the Maxwell construction.
\end{itemize}

I will concentrate my attention on the equilibrium stochastic differential equation. Let us present one of the most striking results: if we consider the leading terms in the perturbative expansion
in the low temperature phase, just below $T_d$, we can consider the following stochastic differential equation:
\be
 -\Delta q_\omega(x) -W'(q_\omega(x))=\omega(x) \, , \label{STOA}
\ee
where $\omega(x)$ is a Gaussian short range noise, (i.e.
$\overline{\omega(x) \omega(y)}= g \delta(x-y)$). The quantity $g$ plays the role of the coupling  constant\footnote{May be it can be better for computing the behaviour in the $\alpha$ region to take a $q$ dependent noise, e.g.  $\overline{\omega(x,q) \omega(y,q')}= g f(q,q') \delta(x-y)$. It is likely that this would be only a minor improvement of the theory. I will not discuss anymore this point.}.

Let us consider the overlap profile at large times. In the region $T<T_d$ we define 
\be
q_\infty(t',x)=\lim_{t\to \infty}q(t',t'+t,x) \ .
\ee
In perturbation theory $q_\infty(t',x)$ is a random (correlated) variable, that depends on the initial conditions because in the region below $T_d$ the dynamics is no more ergodic and the memory of the initial condition is persistent.

We can prove that in perturbation theory the statistical properties of the overlap profile $q_\infty(t',x)$ 
are just the same of the solution of the previous stochastic differential equation, if $g$ is chosen in a suitable way. Therefore one finds for the correlations of the heterogeneities at large time the following relations:
\be
G_0(x) = \overline{q_\omega(x) q_\omega(0)}  -\overline{q_\omega(x)} \;\overline{q_\omega(0)}\, , \ \ \ \ G_1(x) \propto \overline{q_\omega(x) \omega(0)} \, ,
\ee

The physics is quite clear.
The choice of the initial conditions in the dynamics 
induce a fluctuating (point-dependent) linear term in the potential (that induces  local fluctuations of the critical temperature) and the
effects of these fluctuations are the dominant one.

The previous results can be generalized to the late time correlations. Indeed
we can introduce a
fictitious time $s$ and write the following evolution equation:
\be
{\partial q(x,s) \over \partial s} = \Delta q(x,s)
- W'(q(x,s)) + \omega(x)\, ,
\label{FINAL}
\ee
with the boundary conditions at time $s=0$ given by $ q(x,0)=1$.
We can argue that in the whole $\beta$ region (always near to the critical point), if we limit ourselves to reparametrization invariant quantities, the previous equation gives the correct results. It is interesting to note that the previous equation just describes the naive decay of the overlap profile in a system with a potential
\be
\int dx\left( \frac12 (\partial_\mu q(x))^2 -W(q(x))-\omega(x)q(x) \right) \ .
\ee

We remark that the fictitious time $s$ in the previous equation 
should not be identified with the real time: the dependence of $q(s)$,
if we solve equation (\ref{FINAL}), is quite different from the
behaviour of $q(t)$ in mode-coupling theory and there should be no
confusion among the two  time-like variables $s$ and $t$.

\subsection{Beyond Perturbation theory}
It can be shown that eq. ({\ref{STOA}) starts to have many solutions when we approach the dynamical temperature.
In presence of multiple solutions (as it happens for the Random
Field Ising Model) everything becomes more complex (multiple solutions
cannot be seen in perturbation theory so that this problem does not
affect the perturbative analysis that we have presented above). Therefore perturbation theory is ill-defined. This is a problem also for the equilibrium differential equation (\ref{STOA}) because in presence of multiple solutions we have to make a choice and select one of the solutions.

The dynamic formulation, eq. (\ref{FINAL}) is free from these ambiguity and we bet that it gives the correct answer (the question cannot be decided in perturbation theory). Therefore in
the region where the equation has many
solutions, the solution relevant for the dynamics is uniquely
identified by
\be
q^*_\omega(x)=\lim_{s \to \infty} q_\omega(x,s) \ .
\ee

We are near to the end of our journey. We still have to compute the
critical exponent and the critical behaviour of the correlation of
$q^*_h(x)$. This can be easily done in $8-\eps$ dimensions, but in dimension $3$ we fear that only a numerical simulation of equation (\ref{FINAL}) could give the correct results. The simulation is quite easy and there should be no problem in extracting the correct results.

What happens in the more difficult $\alpha$ region? We  conjecture that, \Cred{if we consider only reparametrization
invariant quantities}, the previous stochastic evolution equation gives the correct results for reparametrization invariant quantities, both in the $\beta$ and  in the $\alpha$ region. Indeed the differential equation (\ref{FINAL}) make sense also in the $\alpha$ region: we know already that it gives the correct results in the mean field case, see eq. (\ref{MIRA}).

 \begin{figure}[t!]
  \includegraphics[ width=.7\columnwidth]{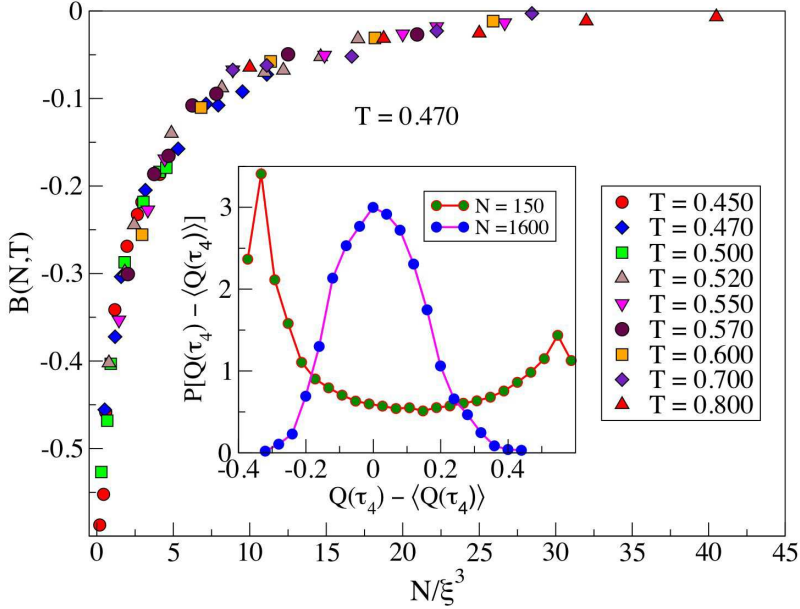}
\caption{ The Binder parameter $\tilde B$ at the time where the susceptibility has a peak as function of $V/\xi^3$ (data taken from \cite{KDS}). Please notice that $\tilde B=-2/3 B$, where $B $ is the conventionally defined Binder parameter. In the insert the probability distribution of the overlap for various $N$. }
\label{B}
\end{figure}

If this conjecture is correct we have a very simple and unified picture for describing the heterogeneities both in the $\beta$ and in the $\alpha$ regions. The behaviour in the $\alpha$
region should be very similar to the one described in the recent paper \cite{KZ}, i.e. the nucleation of regions of small overlap that invade the space.  A hand waving  argument suggest that that in the fully fledged $\alpha$ region, the overlap susceptibility is proportional to $\xi^3$ (as observed firstly  in \cite{PL} at low temperatures and recently \cite{FS} also near the dynamical transition), however numerical simulations are needed to check the validity of the assumptions.

Our approach has many points in common with the one of \cite{KZ}: 
 these authors say that the dynamics of glasses near the dynamical phase transition has many points in common with the process of decay of the unstable phase near the end of the metastable phase. Our point is more radical: when we consider the case where some space disorder is present, the  two physical situations belongs to the same universality class, if  we restrict our study to reparametrization invariant quantities.

Up to now we have not included activated  processes. The previous  conjecture may be naturally generalized by introducing a more general
evolution equation that can be used to compute the properties of
reparametrization invariant quantities when activated processes are
present below $T_c$:
\be
{\partial q(x,s) \over \partial s}= \Delta q(x,s)
-W'(q(x,s))+\omega(x)+\eta(x,s) \label{ACTIVATED}
\ee
where $\eta(x,s)$ is just a thermal noise. Apparently the physics is right. This equation seems to predict that the crossing of the barrier is going through a sequence of noise induced avalanches \cite{EXP}. The previous equation is quite appealing because of its simplicity.
It would be very interesting to carefully compare the predictions of these equations with the experimental data, also in order to understand if something is missing.

We finally consider one of the few presentations of numerical data in a reparametrization invariant way: i.e. the evaluation of the Binder parameter of the overlap in finite volume \cite{FS}( see fig. (\ref{B})) at the point of maximum of the dynamical susceptibility.
 The function $B(V/\xi^3)$ is computable by solving numerically equation (\ref{FINAL}). However we can already see without solving numerically the equation  that there is a qualitative agreement with the theoretical predictions. Indeed in the small volume limit we should get the same result of the mean field theory (see section (\ref{DHMFT}): a bimodal (symmetric) distribution of the overlap and therefore $B=1$, i.e. $\tilde B=-2/3$ in the normalization of figure (\ref{B}). One the other end for large volume we expect a Binder parameter decaying as the inverse of the volume.
 
 It is remarkable that equation (\ref{ACTIVATED}) should also describe the large scale behaviour of the approach to equilibrium in a wide class of systems where avalanches and  crackling noise is present \cite{DS,SDP,WLD}, for example in ferromagnetic systems where quenched randomness is present and Barkhausen noise is present. The subject has been widely studied and the reader should look to the original papers \cite{DS,SDP,WLD}. We only remark that region  that is relevant for the glassy transition is the small randomness region, where one has eventually to take care of the activated processes. Indeed, if activated processes are neglected and we consider equation (\ref{FINAL}) a first order transition is present  \cite{DS,SDP,WLD}.

\section{Conclusions}
We have written various equations, that can be very easily solved numerically, that should be enough to compute the properties of the  dynamical heterogeneities, if we limit ourselves to the sector of reparametrization invariant quantities. 
The results in the $\beta$ region are very solid, while they become less solid when we enter in the $\alpha$ region and we consider activated process.

We have proposed a  single stochastic differential equation that allow us to obtain a unified picture of the heterogeneities related to the glass transition. This equation  also describes  the dynamics near the endpoint of a metastable phase in a disordered system. Therefore, {\em as far as reparametrization invariant quantities are concerned}, heterogeneities in glassy systems have many points in contact with the Barkhausen noise \cite{DS,SDP,WLD} and they should belong to the same universality class.

It would be very interesting to compare the predictions of the theory (including the various critical exponents) with the experimental and numerical data, with the aim of understanding the points of contacts between theory and experiments and to understand the possible discrepancies. In order to reach this goal it would be very important if the (experimental or numerical) data were presented also in reparametrization invariant form, where the time is eliminated in a parametric way. This would help not only the comparison with the present theory, but also among different systems.

\section*{Acknowledgement}
It is a pleasure to thank the organizer of the StatphysHK, Michel Wong and Leihan Tang for providing the occasion to write this paper that profited of the inspiring atmosphere of the conference. We would also thank James Sethna and Stefano Zapperi for very useful discussions on the Barkhausen noise.

\end{document}